\documentclass[aps,prb,preprint,groupedaddress,floatfix]{revtex4}

\usepackage{graphicx,xspace,amsmath}

\newcommand{\eV}{\nobreak\mbox{$\;$eV}\xspace}
\newcommand{\meV}{\nobreak\mbox{$\;$meV}\xspace}
\newcommand{\nm}{\nobreak\mbox{$\;$nm}\xspace}
\newcommand{\GtR}{\!>\!} 
\newcommand{\Wcmsq}{\nobreak\mbox{$\;$W$\,$cm$^{-2}$}\xspace}
\newcommand{\gtR}{>\!} 
\newcommand{\lesS}{<\!} 
\newcommand{\etal}{\nobreak\mbox{\it et al.}\xspace}
\newcommand{\celsius}{\nobreak\mbox{$\;^{\circ}$C}\xspace}
\newcommand{\MinuS}{\!-\!} 
\newcommand{\siM}{\sim\!} 
\newcommand{\um}{\nobreak\mbox{$\;\mu$m}\xspace}
\newcommand{\fs}{\nobreak\mbox{$\;$fs}\xspace}
\newcommand{\kHz}{\nobreak\mbox{$\;$kHz}\xspace}
\newcommand{\sub}[1]{_{\mathrm{#1}}} 
\newcommand{\SiM}{\!\sim\!} 
\newcommand{\invcm}{\nobreak\mbox{$\;$cm$^{-1}$}\xspace}
\newcommand{\cmGW}{\nobreak\mbox{$\;$cm$\,$GW$^{-1}$}\xspace}
\newcommand{\cm}{\nobreak\mbox{$\;$cm}\xspace}
\newcommand{\uJcmsq}{\nobreak\mbox{$\;\mu$J$\,$cm$^{-2}$}\xspace}
\newcommand{\EE}[1]{\!\times\!10^{#1}}
\newcommand{\invvol}{\nobreak\mbox{$\;$cm$^{-3}$}\xspace}
\newcommand{\ns}{\nobreak\mbox{$\;$ns}\xspace}
\newcommand{\ps}{\nobreak\mbox{$\;$ps}\xspace}
\newcommand{\hw}{\hbar\omega}
\newcommand{\LesssiM}{\!\lesssim\!} 
\newcommand{\EQ}{\!=\!} 
\newcommand{\kelvin}{\nobreak\mbox{$\;$K}\xspace}
\newcommand{\spr}[1]{^{\mathrm{#1}}}
\newcommand{\LesS}{\!<\!} 
\newcommand{\LE}{\!\le\!} 
\newcommand{\dg}{\nobreak\mbox{$\,^{\circ}$}\xspace}
\newcommand{\sr}{\nobreak\mbox{$\;$sr}\xspace}
\newcommand{\GtrsiM}{\!\gtrsim\!} 
\newcommand{\PM}{\!\pm\!} 
\newcommand{\EquiV}{\!\equiv\!} 
\newcommand{\ProptO}{\!\propto\!} 
\newcommand{\ApproX}{\!\approx\!} 
\newcommand{\PluS}{\!+\!} 
\newcommand{\lesssiM}{\lesssim\!} 
\newcommand{\e}[1]{e^{#1}} 
\newcommand{\gtrsiM}{\gtrsim\!} 

\begin{document}

\title{Effects of reabsorption and spatial trap distributions\\on the radiative quantum efficiencies of ZnO}

\author{J. V. Foreman$^{\textrm a)}$
\footnotetext{$^{\rm a)}$Electronic mail: john.v.foreman@us.army.mil} }
\affiliation{Department of Physics, Duke University, Durham, North Carolina 27708 and 
U.S. Army Aviation \& Missile Research, Development, \& Engineering Center, Redstone Arsenal, AL 35898 
}

\author{H. O. Everitt}
\affiliation{Department of Physics, Duke University, Durham, North Carolina 27708 and 
U.S. Army Aviation \& Missile Research, Development, \& Engineering Center, Redstone Arsenal, AL 35898 
}

\author{J. Yang}
\affiliation{Department of Chemistry, Duke University, Durham, North Carolina 27708}

\author{T. McNicholas}
\affiliation{Department of Chemistry, Duke University, Durham, North Carolina 27708}

\author{J. Liu}
\affiliation{Department of Chemistry, Duke University, Durham, North Carolina 27708}

\date{June 6, 2009}

\begin{abstract}
Ultrafast time-resolved photoluminescence spectroscopy following one- and two-photon excitation of ZnO powder is used to gain unprecedented insight into the surprisingly high external quantum efficiency of its ``green'' defect emission band.  The role of exciton diffusion, the effects of reabsorption, and the spatial distributions of radiative and nonradiative traps are comparatively elucidated for the ultraviolet excitonic and ``green'' defect emission bands in both unannealed, nanometer-sized ZnO powders and annealed, micrometer-sized ZnO:Zn powders. We find that the primary mechanism limiting quantum efficiency is surface recombination because of the high density of nonradiative surface traps in these powders. It is found that unannealed ZnO has a high density of bulk nonradiative traps as well, but the annealing process reduces the density of these bulk traps while simultaneously creating a high density of green-emitting defects near the particle surface. The data are discussed in the context of a simple rate equation model that accounts for the quantum efficiencies of both emission bands.  The results indicate how defect engineering could improve the efficiency of ultraviolet-excited ZnO:Zn-based white light phosphors.
\end{abstract}

\pacs{78.47.Cd, 78.55.Et, 78.67.Bf}

\maketitle


\section{\label{secIntro} Introduction}

In the current renaissance~\cite{Klingshirn07,Klingshirn072} of research related to zinc oxide (ZnO), most attention is focused on the material's near-band-edge optical properties for the development of new/improved ultraviolet (UV) emitters and detectors.  In addition to its direct, wide band gap of $3.4\eV$, an exciton binding energy of $60\meV$ is frequently cited as a primary motivation to investigate ZnO.  However, ZnO has also proven viable as a green/monochrome phosphor in electron-excited~\cite{Pfahnl62,Vecht94,Holloway99} and photon-excited~\cite{Foreman06} applications. This highly efficient ``green band'' luminescence typically peaks around $2.5\eV$ ($500\nm$) and essentially spans the entire visible spectrum in a manner well matched to the human eye response. The exponentially growing market for inexpensive, environmentally friendly solid-state lighting has made ZnO-based phosphors particularly compelling. Although the general consensus is that a lattice defect (possibly an oxygen or zinc vacancy) rather than an impurity is responsible for green band luminescence,~\cite{Ozgur05} ZnO lattice imperfections in general and the microscopic origin of ZnO green emission in particular remain a matter of debate. Resolving the influence of these defects on the optical properties of ZnO will allow intentional engineering of ZnO for either UV or green/white optoelectronic applications. 

Toward that end, we have previously reported on the remarkable enhancement of green band emission from doped ZnO powders and nanostructures following UV excitation.~\cite{Foreman06, Foreman07}  In those investigations, it was found that the emission strength dramatically increased after sulfur doping and annealing but decreased with increasing excitation intensity $\GtR0.2\Wcmsq$.  It was also found that the external quantum efficiency dropped from $\gtR65\%$ for micron-sized powders to $\lesS30\%$ for nanometer-sized wires, a confirmation that nanostructuring increases the probability of competing nonradiative relaxation.~\cite{Foreman06} Although qualitatively compelling, the model\cite{Shalish04} proposed by Shalish \etal underestimated the amount of green defect emission from these sulfur-doped, annealed ZnO nanostructures by more than an order of magnitude.  Much work remains to understand the relationship between defects and optical performance.

Here we introduce a new experimental tool---namely comparative ultrafast one- and two-photon excitation spectroscopy---that, when combined with traditional photoluminescence (PL), photoluminescence excitation (PLE) spectroscopy, and quantum efficiency (QE) measurements, provides a more complete and quantitative understanding of the relationship between excitonic and defect emission with competing nonradiative recombination processes. Because one-photon excitation creates excitons near the particle's surface while two-photon excitation creates excitons throughout the particle's volume, this technique provides unprecedented insights into exciton diffusion, the effects of reabsorption, and the spatial distributions of radiative and nonradiative traps. We use this technique to compare the PL spectra and recombination lifetimes in unannealed, nanometer-sized ZnO powders and annealed, micrometer-sized ZnO:Zn powders.  Unlike sulfur-doped ZnO, the UV and green defect band emission features are well-resolved in these materials, allowing each emission process to be investigated separately. One- and two-photon excitation is used to measure the external quantum efficiency, systematically quantifying how the UV and green emission strengths depend on the spatial distribution and lifetimes of radiative and nonradiative traps. These experimental investigations indicate a competition among three excitonic relaxation pathways:  Radiative recombination, nonradiative recombination, and transfer to a green defect emission site.  This competition is reflected in a simple rate equation model that uses the measured relaxation rates to predict the quantum efficiencies and suggest how defect engineering could improve them.

\section{\label{secExpt} Experimental Procedures}
The ZnO samples were derived from commercially available ZnO powder (99.999\%, Alfa Aesar).  A reference sample, hereafter referred to as ``unannealed ZnO,'' underwent no additional treatment.  A second sample (``annealed ZnO'') was prepared by sealing a portion of the unannealed ZnO in a quartz tube under vacuum and firing it for one hour at 1000\celsius in a horizontal tube furnace. This procedure is similar to the commercial production of ``zinc-doped ZnO,'' or ZnO:Zn (to denote an excess of zinc atoms or a deficiency of oxygen atoms), where ZnO powder is annealed between $900\MinuS1100\celsius$.~\cite{Vesker81,Yen04}

\begin{figure}[] \begin{centering}
\includegraphics[width=8cm]{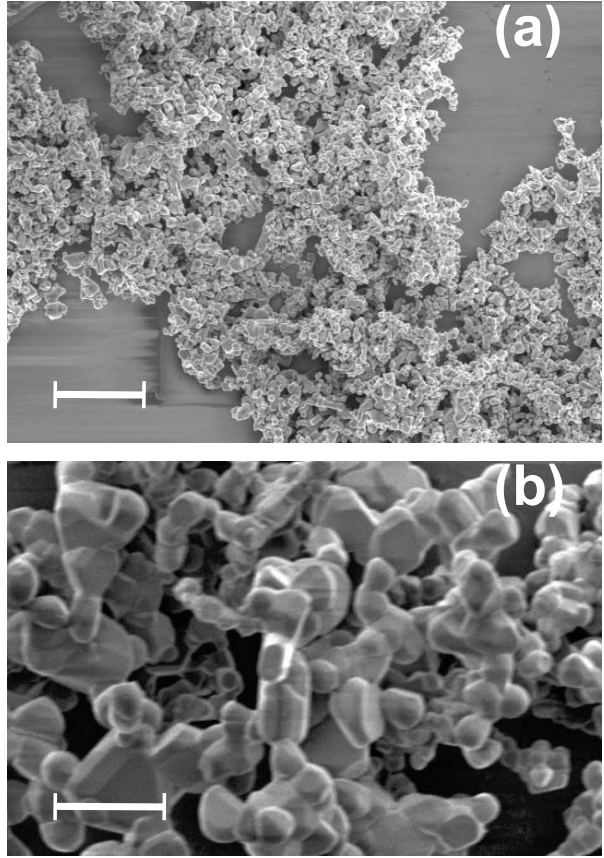}
\caption{\label{figPowderSEM} SEM images of (a) unannealed ZnO powder and (b) annealed ZnO powder.  Both scale bars correspond to $5\um$.}
\end{centering} \end{figure}

SEM images (Fig.~\ref{figPowderSEM}) indicate that the unannealed powder particles have a median diameter of 340\nm, whereas the annealed powder exhibits a much larger median size of $1800\nm$.  As discussed in Ref.~\onlinecite{Foreman07}, the annealed ZnO sample exhibits optical properties virtually identical to those of typical ZnO:Zn phosphor powders; i.e., the photoluminescence spectrum is dominated by highly efficient, defect-related emission centered at $\siM 2.5\eV$ with a full width at half-maximum of $\siM 450\meV$.  In the unannealed ZnO sample, this ``green band'' defect emission is almost three orders of magnitude weaker than the ultraviolet, near-band-edge emission.

The samples were prepared for optical characterization by mixing them with methanol, dropping the mixtures onto microscope slides, and allowing the methanol to evaporate.~\cite{FNsolvent}  Optical excitation was provided by $\siM 100\fs$ pulses from a tunable 1\kHz optical parametric amplifier (OPA).  Above-gap excitation at 3.84\eV (323\nm) was generated using the fourth harmonic of the OPA's signal.  Due to the large absorption coefficient of ZnO at this excitation energy ($\alpha\sub{exc} \SiM 2\times 10^5 \invcm$), creation of electron-hole pairs occurs primarily near the surface of the particles ($\alpha^{-1}\sub{exc} \SiM 50\nm$).~\cite{Yoshikawa97,Jellison98}  The samples were also excited below the band gap at 1.92\eV (646\nm) using the second harmonic of the OPA's signal.  In this case, electron-hole pairs are generated via two-photon absorption, but the carriers have the same initial energy distribution as for one-photon excitation since the two-photon excitation energy is half that of the one-photon excitation.  Assuming a two-photon absorption coefficient of  $\beta\sub{exc} \SiM 3 \cmGW$ for ZnO,~\cite{Bolger93,He05} the effective penetration depth of two-photon excitation is $\siM 1\cm$; thus, the carriers are generated uniformly throughout the volume of the nanometer- and micrometer-scale particles.  The incident laser fluences were approximately $3\uJcmsq$ and $80\uJcmsq$ for one-photon (``surface'') and two-photon (``volume'') excitation, respectively.  After taking into consideration the actual number of photons absorbed by the samples (96\% and 8\% for one-photon and two-photon excitation, respectively~\cite{FNabsorb}) and the fact that two photons of volume excitation lead to a single photoexcited electron-hole pair, the effective fluence for both one-photon and two-photon excitation is $\siM 3\uJcmsq$.  This fluence corresponds to a photogenerated carrier density of approximately $1\EE{18}\invvol$.
	
Time-integrated photoluminescence (TIPL) spectra were measured using a pair of off-axis parabolic mirrors to collect and focus the PL into a fused silica optical fiber that was coupled to a 30\cm spectrometer with CCD detector (Acton/Princeton Instruments).  Time-resolved PL (TRPL) was measured using the same collection optics in conjunction with an analog streak camera (Hamamatsu C4334).  The streak camera's time window and resolution were 10\ns and 140\ps for acquisition of near-band-edge emission.  External quantum efficiency measurements were made under the same excitation conditions using a 10\cm diameter integrating sphere, as described previously.~\cite{deMello97}  All PL spectra were corrected for the spectral sensitivities of the various optical paths and detectors, and unless otherwise specified, all measurements were performed at room temperature.

\section{\label{secResults} Results and Discussion}
%
\subsection{\label{secReabsorption} One- versus two-photon excitation: how reabsorption, reflections, and trap distributions affect quantum efficiency}
%

After photoexcitation of an exciton population in ZnO, luminescence is emitted at energies $\hw\LesssiM E\sub{X}$, where $E\sub{X}\SiM3.3\eV$ is the exciton energy. Free exciton luminescence occurs at $\hw\SiM E\sub{X}$, and lower-energy emission occurs through longitudinal optical (LO) phonon-assisted exciton annihilation (hereafter denoted X-$m$LO for involvement of $m\EQ0,1,2,\ldots$ LO phonons).~\cite{Klingshirn05,Klingshirn07}  Depending on their energy, these luminescence photons are subject to reabsorption through various mechanisms (band-to-band absorption, exciton resonance, below-gap ``Urbach'' tail absorption, etc.).  The probability of reabsorption is described by the strength of the absorption spectrum $\alpha(\hw)$.  On average the photon will reach the sample surface without undergoing reabsorption if it originates within the optical escape depth $\alpha^{-1}(\hw)$ of the surface.  Whenever a reabsorption event does occur, a free electron-hole pair is re-created.  This electron-hole pair is then subject anew to the possibilities of exciton formation and/or energy transfer to ``traps'' (either radiative or nonradiative).

Quantifying $\alpha(\hw)$ for the powder samples in this work is difficult because the samples are optically thick and highly scattering.  We have adopted a technique based on PL excitation (PLE) spectroscopy to estimate the absorption spectrum.  The underlying assumption of this technique is that the PLE spectrum of near-band-edge luminescence---specifically the X-2LO transition---effectively maps out the exciton and band-to-band absorption resonances because many combinations of phonon wave vectors satisfy the momentum conservation rule associated with 2-LO-assisted exciton annihilation. Therefore, the properties of this luminescence feature are representative of the entire photoexcited exciton population.~\cite{Klingshirn05,Permogorov82}

\begin{figure}[]
\begin{center}
\includegraphics[width=8cm]{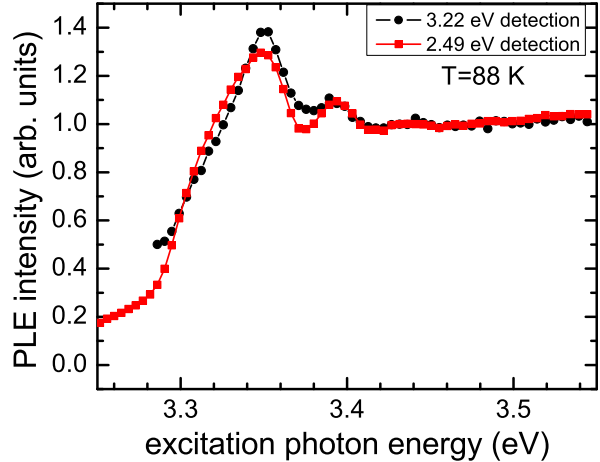}
\end{center}
\caption{\label{figLTPLE} (Color online) Comparison of photoluminescence excitation spectra for detection energies near the peak of the green band ($2.49\eV$) and at the peak of the X-2LO transition ($3.22\eV$) for the annealed ZnO:Zn powder sample.  The lattice temperature was $88\kelvin$.}
\end{figure}

Figure~\ref{figLTPLE} shows PLE spectra for detection wavelengths near the peak of the green PL (2.49 eV) and at the X-2LO transition (3.22 eV) for the annealed ZnO sample.  The lattice temperature was $T\EQ88\kelvin$, and the two curves have been normalized to their intensities at the band gap ($E\sub{g}\EQ3.43\eV$).~\cite{FNltple}  The excitation profiles for exciton luminescence and green luminescence are almost identical.  In addition to suggesting that efficient energy transfer to green-emitting defects depends directly on a pre-existing exciton population, this result suggests that the \textit{green band} PLE can also be used to estimate $\alpha(\hw)$.~\cite{Wang03}

\begin{figure}[]
\begin{center}
\includegraphics[width=8cm]{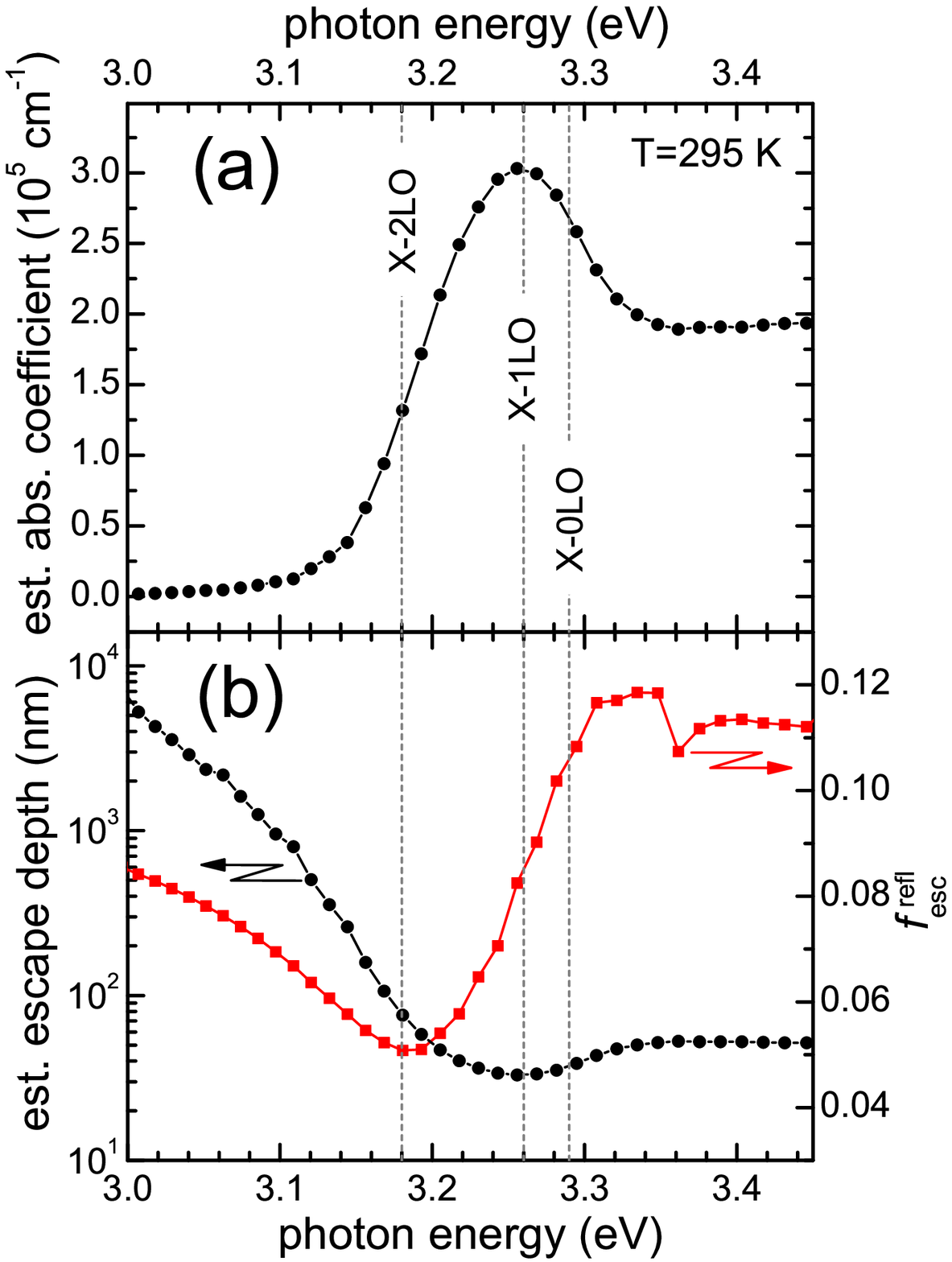}
\end{center}
\caption{\label{figPLEopticalEscape} (Color online) (a) Estimated absorption coefficient of ZnO:Zn powder as derived from the photoluminescence excitation spectrum of the peak of the green defect emission.  (b) Estimated photon absorption or escape depth [inverse of the curve in (a)] due to reabsorption, and the probability $f\sub{esc}\spr{refl}$ of photon escape due to Fresnel reflection [derived from Eq.~\eqref{eqFresnelTrans} and a Kramers-Kronig transformation of the curve in (a)].}
\end{figure}

The room-temperature absorption spectrum for the annealed ZnO sample was approximated by measuring its green PLE spectrum, then matching the above-band gap ($\hw\EQ3.5\MinuS3.8\eV$) PLE intensity with the typical absorption coefficient of $\alpha\SiM2\times10^5\invcm$ (Fig.~\ref{figPLEopticalEscape}(a)).~\cite{Muth99,Liang68}  The enhancement in absorption at $3.26\eV$, which is below the exciton resonance $E\sub{X}\EQ3.31\eV$,~\cite{FNrtple} was observed previously in the PLE of ZnO nanorods and was attributed to enhanced exciton-phonon coupling at room temperature.~\cite{He07} The energy-dependent photon penetration (or escape) depth $\alpha^{-1}(\hw)$ may be estimated from the inverse of the absorption spectrum in Fig.~\ref{figPLEopticalEscape}(a) and is plotted in Fig.~\ref{figPLEopticalEscape}(b).  

The insight provided by comparing PL and TRPL upon one- and two-photon excitation arises from the differing absorption depths and corresponding photoexcited exciton distributions. Because of the large absorption coefficient of ZnO for $\hw\GtR E\sub{g}$, one-photon excitation primarily generates carriers (and therefore excitons) within $50\MinuS70\nm$ of the particle surface. Two-photon excitation with $(E\sub{g}/2)\LesS\hw\LesS E\sub{g}$ generates the same number of carriers, but over the entire thickness of the sample.  In this case, many excitons will form within the sample at distances exceeding the optical escape depth.  

Now, exciton polaritons with $\hw\SiM3.15\MinuS3.3\eV$ propagate through the sample and undergo one reabsorption event every $\alpha^{-1}(E\sub{X})\SiM50\nm$ on average.  The only exciton polaritons that have a high probability of escaping as luminescence photons are those generated within $\lesssiM50\nm$ of the particle surface.  The rest will mostly be reabsorbed and lost through nonradiative recombination in the bulk of the sample.  Excitons which undergo phonon-assisted recombination in the bulk of the material will yield photons experiencing reabsorption every $\alpha^{-1}(\hw\sub{X-1LO})\MinuS\alpha^{-1}(\hw\sub{X-2LO})\EQ35\MinuS80\nm$.  The probability of escape is higher for the X-2LO transition ($\hw\sub{X-2LO}\EQ3.18\eV$) compared to the X-1LO  and X-0LO transitions ($\hw\sub{X-1LO}\EQ3.26\eV$ and $\hw\sub{X-0LO}\EQ3.29\eV$, respectively) because the escape depth is larger.~\cite{FNphononpeaks}  But again, the most likely fate for electron-hole pairs formed by reabsorption is nonradiative recombination.

Differences in the one- and two-photon PL and TRPL signals arise from the different recombination processes active in each spatial distribution of excitons.  For one-photon/surface excitation, the effective carrier concentration is reduced at the surface due to nonradiative surface recombination.~\cite{Bebb72,Bukesov02,Bylander78,Yoo97}  Therefore, for one-photon excitation we expect the near-band-edge luminescence to originate near the particle surface, to undergo minimal reabsorption, and to experience competition primarily from nonradiative \textit{surface} recombination.  For two-photon/volume excitation, the measured UV luminescence signal will consist primarily of luminescence from the small fraction of excitons photogenerated near the surface of the sample if the density of \textit{bulk} nonradiative traps is significant. Because the absorption coefficient is negligible at visible wavelengths, the green PL suffers no reabsorption and should be independent of the excitation distribution (one-photon versus two-photon).

When the PL generated at a particular photon energy $\hw$ propagates through the material and reaches the ZnO--air interface, it will be totally internally reflected if the angle of incidence $\theta$ is larger than the critical angle $\theta\sub{c}(\hw)\EQ\sin^{-1}[1/n(\hw)]$.~\cite{Jackson99}  Light incident at angles $\theta\LE\theta\sub{c}$ comprises a cone of luminescence that escapes, the average transmittance of which may be estimated by integrating the relevant Fresnel equations:
\begin{align}\label{eqFresnelTrans}
f\sub{esc}\spr{refl}(\hw)=\left(\frac{1}{2\pi}\right)\int_0^{2\pi}\!\int_0^{\theta\sub{c}(\hw)} \Bigg\{&\frac{\sqrt{1-n^2\sin^2\theta}}{n\cos\theta} \notag\\
 \times&\left[ t(\theta,\hw) \right]^2  \Bigg\} \sin\theta \,d\theta d\phi \, ,
\end{align}
where $t(\theta,\hw)$ is the angle- and polarization-dependent Fresnel equation for transmission of photons with energy $\hw$ at a dielectric--air interface.~\cite{Hecht,FNnormalization}

The escape fraction $f\sub{esc}\spr{refl}(\hw)$ averaged over both polarizations of emitted light is shown in Fig.~\ref{figPLEopticalEscape}(b).  The fraction was calculated according to Eq.~\eqref{eqFresnelTrans} using an index of refraction $n(\hw)$ derived from a Kramers-Kronig transformation of the estimated absorption spectrum in Fig.~\ref{figPLEopticalEscape}(a).  Total internal reflection plays a dominant role in reducing the measurable UV PL:  An index of refraction in the range $n\EQ2.8\MinuS2.0$ corresponds to a range of small critical angles $\theta\sub{c}\EQ21\dg\MinuS30\dg$.  The solid angle of the cone of luminescence is therefore $\Omega\sub{c}\EQ2\pi(1\MinuS\cos\theta\sub{c})\EQ0.42\MinuS0.84\sr$ and, when normalized to the hemisphere solid angle $\Omega\sub{h}\EQ2\pi\sr$, corresponds to an escape probability $\Omega\sub{c}/\Omega\sub{h}\EQ0.066\MinuS0.13$.  The small emission cone is therefore largely responsible for the small values of $f\sub{esc}\spr{refl}\LesssiM 12\%$ shown in Fig.~\ref{figPLEopticalEscape}(b).

If, subsequent to reabsorption, nonradiative recombination is probable (due to a high density of traps), then the escape fraction $f\sub{esc}\spr{refl}$ is a reasonable upper limit on the near-band-edge quantum efficiency.  On the other hand, green luminescence photons with $\hw\LesS3\eV$ will not be reabsorbed subsequent to internal reflection; they will continually propagate through the material and reflect from the ZnO--air interface until sucessfully escaping the material.  Fresnel reflections and reabsorption therefore do not place an upper limit on the quantum efficiency of green luminescence.

\subsection{\label{secUnannealed} UV photoluminescence of unannealed ZnO: the distribution of nonradiative traps}
%

\begin{figure}[]
\begin{center}
\includegraphics[width=8cm]{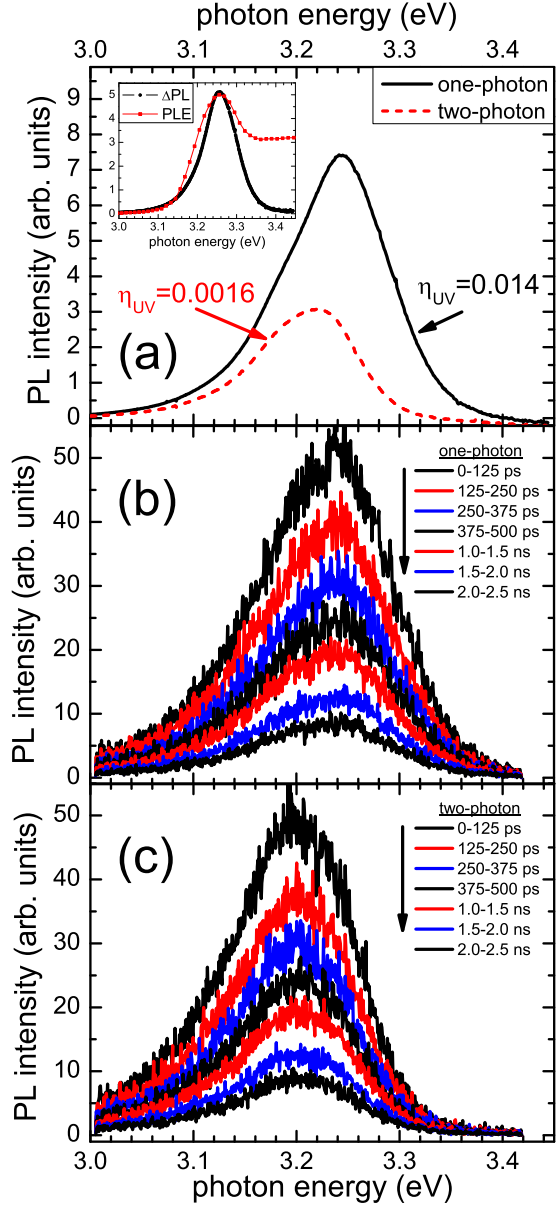}
\end{center}
\caption{\label{figSRTRPLuntrt} (Color online) (a) Time-integrated photoluminescence (TIPL) of unannealed ZnO under one-photon and two-photon excitation.  Time-resolved photoluminescence spectra are also shown for one-photon excitation (b) and two-photon excitation (c).  The inset of (a) demonstrates the effect of reabsorption by comparing the difference in one-photon and two-photon TIPL intensities ($\Delta$PL) to the approximate absorption coefficient derived from photoluminescence excitation (PLE) data.  In (b) and (c), the bottom three spectra in each plot are multiplied by a factor of two for clarity.}
\end{figure}

We now compare the time-integrated and time-resolved PL spectra for surface (one-photon) and volume (two-photon) excitation of the unannealed ZnO powder to ascertain the spatial distribution of nonradiative recombination centers.  The time-integrated PL (TIPL) spectra shown in Fig.~\ref{figSRTRPLuntrt}(a) are as-measured and have not been normalized.  Near-band-edge emission consists of a superposition of free exciton luminescence and its longitudinal-optical (LO) phonon replicas.~\cite{Klingshirn07}

For surface excitation of the unannealed ZnO, reabsorption effects are minimal, as evidenced by the fact that the luminescence extends non-negligibly to energies above the exciton resonance ($E\sub{X}\EQ3.31\eV$).  The high-energy tail of the X-0LO line shape yields a reasonable effective temperature of $T\sub{X}\SiM330\kelvin$ for the Boltzmann-like exciton population.~\cite{Bebb72}  Significant reabsorption at the exciton resonance would increase the slope of the high-energy tail and yield an effective exciton temperature that is unrealistically low (less than room temperature).  The spectrally integrated external quantum efficiency (QE) $\eta\sub{UV}$ for near-band-edge emission, also denoted in Fig.~\ref{figSRTRPLuntrt}(a), is highest for this sample and excitation condition.  However, the low value of $\eta\sub{UV}\EQ1.4\%$ (consistent with other reported values for ZnO\cite{Foreman07,Klingshirn05}) indicates the overall dominance of nonradiative recombination in the ZnO nanoparticles.  Because the majority of carriers are excited within $\alpha^{-1}\sub{exc}\SiM50\nm$ of the particle surface, nonradiative \textit{surface} recombination is expected to play a particularly significant role in depleting carriers and reducing the QE.

In comparing the PL spectra for surface and volume excitation of the unannealed ZnO, it is clear from Fig.~\ref{figSRTRPLuntrt}(a) that the high-energy side of the spectrum is suppressed for volume excitation; furthermore, $\eta\sub{UV}$ is lower by an order of magnitude.  For volume excitation, reabsorption of luminescence is more likely because most photogenerated exciton polaritons must travel a longer distance to reach the surface of the particle.  Following reabsorption, the probability of nonradiative recombination is proportional to the density of bulk nonradiative centers.  For the unannealed sample, the density of bulk nonradiative traps throughout the volume of the particle is evidently significant because the QE drops so dramatically compared to the case of surface excitation.

While one would expect reabsorption to be strongest at energies $\hw\GtR E\sub{X}$, the estimated absorption coefficient trace of Fig.~\ref{figPLEopticalEscape}(a) confirms that the Urbach tail of the absorption is enhanced by the resonance near the X-1L0 energy ($\hw\sub{X-1LO}\EQ3.26\eV$) at room temperature.  Thus, for volume excitation the suppression of the high-energy edge of the UV PL spectrum for $\hw\GtrsiM(E\sub{X} - E\sub{LO})$ and the overall reduction in UV QE is caused by reabsorption and subsequent nonradiative recombination by a high density of \textit{bulk} traps.  This reabsorption effect is dramatically illustrated in the inset of Fig.~\ref{figSRTRPLuntrt}(a), where the difference in one-photon and two-photon PL intensities is virtually identical to the sub-band-gap absorption resonance derived from photoluminescence excitation data [Fig.~\ref{figPLEopticalEscape}(a)].

\begin{figure}[]
\begin{center}
\includegraphics[width=8cm]{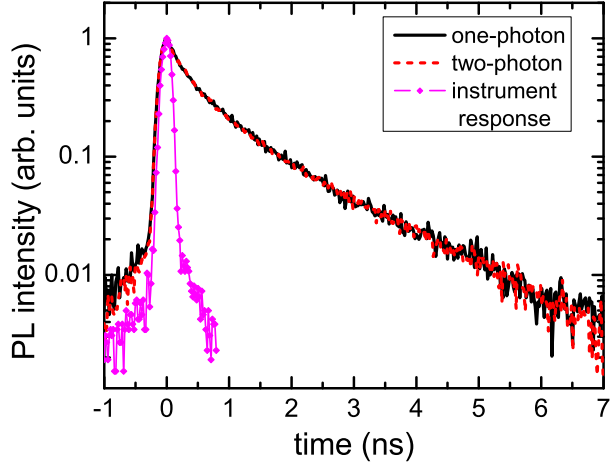}
\end{center}
\caption{\label{figTRPLunannealed} (Color online) Time-resolved photoluminescence of unannealed ZnO under one-photon and two-photon excitation.  The corresponding window of spectral integration was centered at the X-2LO transition ($3.177\eV\PM0.036\eV$).}
\end{figure}

Figure~\ref{figTRPLunannealed} shows the normalized decay of the unannealed ZnO PL spectrally integrated about the X-2LO transition ($3.177\eV\pm0.036\eV$), which is representative of the overall exciton population decay.  There is no significant change in the PL dynamics for surface versus volume excitation because in both cases the collected PL signal originates within the optical escape depth near the particle surface ($\alpha^{-1}\SiM80\nm$ at the X-2LO transition energy).  What \textit{is} different is the number of excitons that either are photoexcited near or have propagated to the surface region of the sample where photon escape is possible, as indicated by the different QE values and the relative PL intensities.

The PL decays of Fig.~\ref{figTRPLunannealed} are well described by a biexponential function, $I\sub{X}(t)\EQ A\sub{f} \exp(-t/\tau\sub{f})+A\sub{s} \exp(-t/\tau\sub{s})$, where $A\sub{f}+A\sub{s}\EQ1$.  The best-fit decay parameters, obtained by taking into account the streak system response function, are shown in Table~\ref{tabUVparams}.  For both excitations of unannealed ZnO the decay is identical:  The fast component of the PL decays with $\tau\sub{f}\SiM400\ps$ and relative amplitude $A\sub{f}\SiM0.7$, while the slow component decays with $\tau\sub{s}\SiM1.5\ns$.

The fast decay component is comparable to what has been measured for bulk single-crystal ZnO ($\tau\sub{f}\spr{bulk}\SiM430\ps$) and has been associated with predominantly nonradiative recombination.~\cite{Teke04}  Given the shallow escape depth of near-band-edge luminescence in ZnO, the similarity of fast decay times for bulk single-crystal and nanometer-scale samples suggests more specifically that nonradiative \textit{surface} recombination competes heavily with exciton-related luminescence during the first few hundred picoseconds following femtosecond excitation.  The fast decay component is much more dominant for ZnO nanoparticles ($A\sub{f}\SiM0.7$) compared to bulk single-crystal samples ($A\sub{f}\spr{bulk}\SiM0.3$) because the surface-to-volume ratio is much larger.

\begin{table}[tbp]
\caption{Biexponential decay best-fit parameters for the X-2LO transition in ZnO particles.  Parameters for bulk single-crystal ZnO (Ref.~\onlinecite{Teke04}) are shown for comparison.} 
\label{tabUVparams}
\begin{center}       
\begin{tabular}{lccc}
\hline
\hline
\rule[-1ex]{0pt}{3.5ex}  Sample/Excitation & $\tau\sub{f}$ [ns] & $\tau\sub{s}$ [ns] & $A\sub{f}$ \\
\hline
\rule[-1ex]{0pt}{3.5ex}  Unannealed/one-photon & 0.40 & 1.6 & 0.72 \\
\rule[-1ex]{0pt}{3.5ex}  Unannealed/two-photon & 0.38 & 1.5 & 0.69 \\
\rule[-1ex]{0pt}{3.5ex}  Annealed/one-photon & 0.33 & 1.7 & 0.77 \\
\rule[-1ex]{0pt}{3.5ex}  Annealed/two-photon & 0.45 & 2.1 & 0.47 \\
\rule[-1ex]{0pt}{3.5ex}  Bulk/one-photon\cite{Teke04} & 0.43 & 3.0 & 0.29 \\
\hline
\hline
\end{tabular}
\end{center}
\end{table}

The slow decay component for bulk single-crystal ZnO is characterized by $\tau\sub{s}\spr{bulk}\SiM3\ns$ (Table~\ref{tabUVparams}) and has been attributed to the intrinsic radiative lifetime of the free exciton.~\cite{Teke04}  The reduction in $\tau\sub{s}$ by a factor of $\siM2$ for nanoparticles is likely due to increased competition for carriers within the volume of the material because of a higher density of \textit{bulk} nonradiative recombination centers.

Associating the fast/slow PL decay components with recombination at nonradiative surface/bulk traps is supported by an analysis of the time-resolved PL \textit{spectra} [Fig.~\ref{figSRTRPLuntrt}(b)--(c)]. Neither spectral distribution exhibits a shift in energy as a function of time, and this has significant consequences that restrict possible decay mechanisms.

Absence of a time-dependent energy shift in surface-excited PL is inconsistent with trapped excitons as the slow decaying component.~\cite{Herz99}  If exciton trapping were occurring, then the spectrum would redshift as a function of time:  At very early times, the PL would primarily correspond to free excitons that have not yet become trapped.  After a delay characterized by the exciton localization time, the PL would primarily correspond to trapped excitons which, by definition, would emit photons of lower energy upon annihilation.  Moreover, if exciton trapping were responsible for the biexponential decay, then the decay characteristics would depend sensitively on the relative densities of localized states and nonradiative surface traps.  Generally speaking, as the sample's surface-to-volume ratio increases, the relative density of nonradiative surface traps would increase and the fast decay component (associated with free exciton plus nonradiative surface recombination) would become more dominant than the slow component (associated with localized exciton recombination).  However, because the fast decay amplitudes for both unannealed and annealed ZnO are $A\sub{f}\SiM0.75$ (Table~\ref{tabUVparams}) while the samples' surface areas differ by a factor of $\SiM30$, it is unlikely that exciton trapping is responsible for the observed biexponential decay.

Absence of a time-dependent energy shift in the surface-excited PL also eliminates the possibility that the biexponential decay of the PL is explained by exciton diffusion from the surface region to the bulk region.  Compared to volume excitation, where the excitons are generated uniformly throughout the particle, surface excitation creates a strong gradient in the spatial distribution of excitons.  One might expect the excitons to diffuse away from the surface while simultaneously undergoing nonradiative surface recombination and emitting exciton luminescence.  These processes would presumably occur on a fast time scale and would correspond to the fast component of the PL decay, whereas the slow component would correspond to relaxation once the excitons have diffused and equilibrated throughout the volume of the particle.  If this type of diffusion were occurring in our sample, then the PL spectrum would redshift during the decay; it would evolve from the spectrum observed for surface excitation [the one-photon curve in Fig.~\ref{figSRTRPLuntrt}(a)] to the spectrum observed for volume excitation [the two-photon curve in Fig.~\ref{figSRTRPLuntrt}(a)].  However, the spectra plotted in Fig.~\ref{figSRTRPLuntrt}(b) show that this is not the case.  Likewise, the spectra in Fig.~\ref{figSRTRPLuntrt}(c) indicate that when the excitons are uniformly generated throughout the sample, there is no net migration of excitons toward the surface as a function of time.  If this were the case, then the two-photon-excited PL would blueshift during the decay from the spectrum observed for volume excitation to the spectrum observed for surface excitation.

Because neither spectral distribution shifts in energy during the course of PL decay, there is no large-scale spatial redistribution of excitons occurring via surface--bulk or bulk--surface diffusion.  The data indicate that the distribution of excitons in the sample is relatively unchanged from the moment of photoexcitation to the completion of the decay process, regardless of whether surface or volume excitation is used.  The time-resolved PL spectra thus support the claims that 1) the differences in time-integrated PL spectra and QEs are due to differences in initial spatial distributions of carriers, 2) the extent of reabsorption and bulk nonradiative recombination is much more significant for volume excitation, and 3) the biexponential decays are explained by distinct surface and bulk nonradiative recombination mechanisms rather than exciton trapping and/or surface--bulk diffusion.

\subsection{\label{secAnnealedUV} UV photoluminescence of annealed ZnO: the distribution of green-emitting defects}
%

\begin{figure}[]
\begin{center}
\includegraphics[width=8cm]{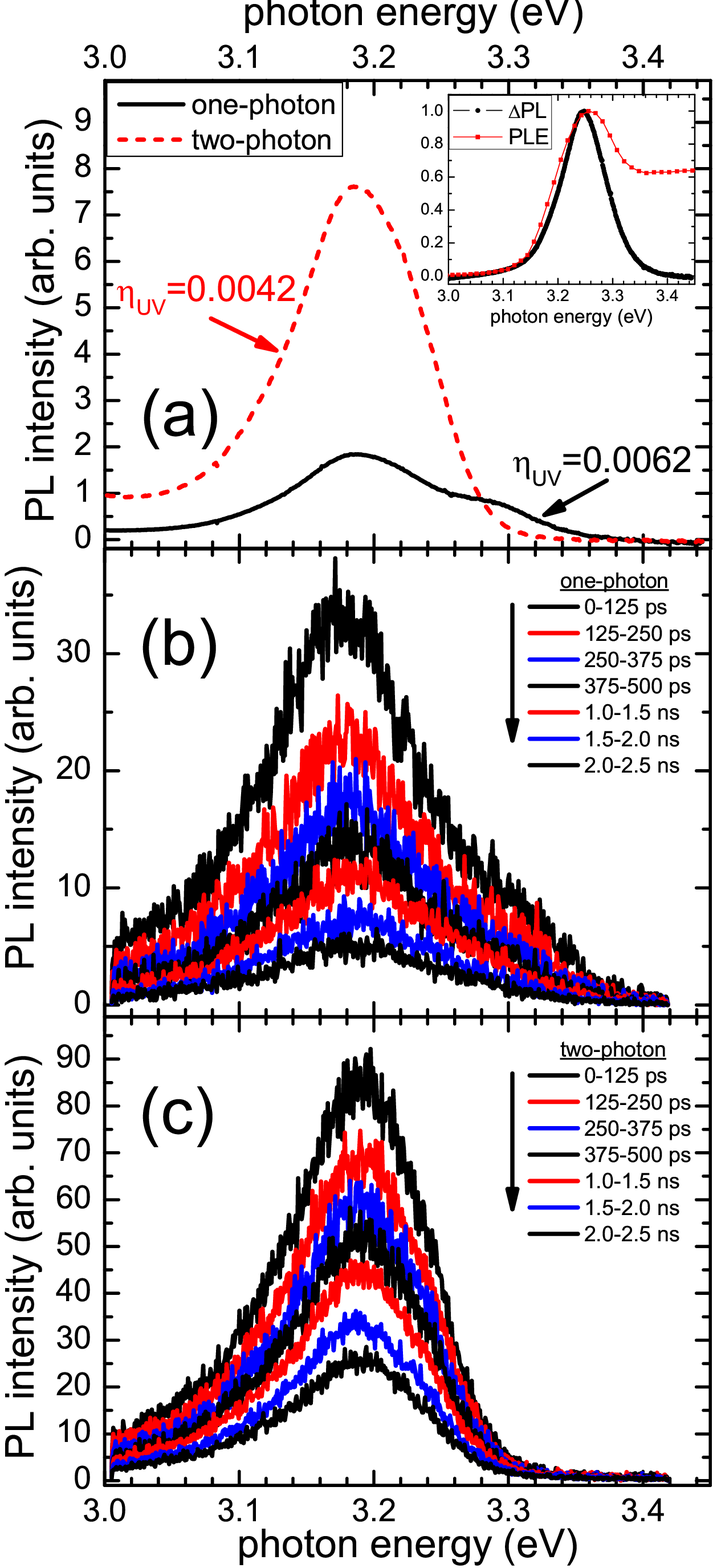}
\end{center}
\caption{\label{figSRTRPLtrt} (Color online) (a) Time-integrated photoluminescence (TIPL) of annealed ZnO under one-photon and two-photon excitation.  Time-resolved photoluminescence spectra are also shown for one-photon excitation (b) and two-photon excitation (c).  The inset of (a) demonstrates the effect of reabsorption by comparing the difference in one-photon TIPL intensities ($\Delta \mathrm{PL}\EQ \mathrm{PL}\sub{unannealed}-\mathrm{PL}\sub{annealed}$) to the approximate absorption coefficient derived from photoluminescence excitation (PLE) data.  In (b) and (c), the bottom three spectra in each plot are multiplied by a factor of two for clarity.}
\end{figure}

The near-band-edge TIPL spectra for the annealed ZnO sample are shown in Fig.~\ref{figSRTRPLtrt}(a).  As described previously,~\cite{Foreman07} the annealing process creates a large number of green-emitting defects.  The presence of these defects has a pronounced effect on the near-band-edge PL when the annealed sample is excited primarily near the surface.  The inflection point at $\siM3.26\eV$ in the one-photon PL spectrum [Fig.~\ref{figSRTRPLtrt}(a)] lies near the peak of the effective absorption coefficient derived from the green defect excitation profile [Fig.~\ref{figPLEopticalEscape}(a)], suggesting that the change in PL compared to the case of unannealed ZnO is due to additional reabsorption caused by the presence of green-emitting defects.  The spatial distribution of these defects is suggested by the inset of Fig.~\ref{figSRTRPLtrt}(a), which compares the difference in surface-excited (one-photon) photoluminescence: $\Delta \mathrm{PL}\EQ \mathrm{PL}\sub{unannealed}-\mathrm{PL}\sub{annealed}$. The $\Delta$PL is very similar in shape to the sub-band-gap absorption resonance in the green PLE spectrum, indicating that additional reabsorption is indeed occurring \textit{near the surface} of the annealed ZnO sample.

When the entire volume of the annealed particle is excited via two-photon excitation, the blue side of the PL spectrum is again suppressed relative to the case of one-photon excitation due to reabsorption [Fig.~\ref{figSRTRPLtrt}(a)].  However, the annealing process appears to reduce the density of bulk nonradiative traps because the two-photon $\eta\sub{UV}$ does not drop as drastically from its one-photon value (from 0.62\% to 0.42\%) as it did in the case of unannealed ZnO (from 1.4\% to 0.16\%), despite the additional presence of ultraviolet-absorbing, green-emitting defects.  Thus, reabsorption of volume-excited excitons still occurs---both by bulk nonradiative traps and by green-emitting defects.  In the former case, the probability of subsequent nonradiative recombination has been reduced because the annealing process has reduced the density of bulk traps (evidenced by a higher volume-excited $\eta\sub{UV}$).  In the latter case, subsequent emission of a green photon is highly probable given the high QE of green-emitting defects (Sec.~\ref{secAnnealedGR}).

\begin{figure}[]
\begin{center}
\includegraphics[width=8cm]{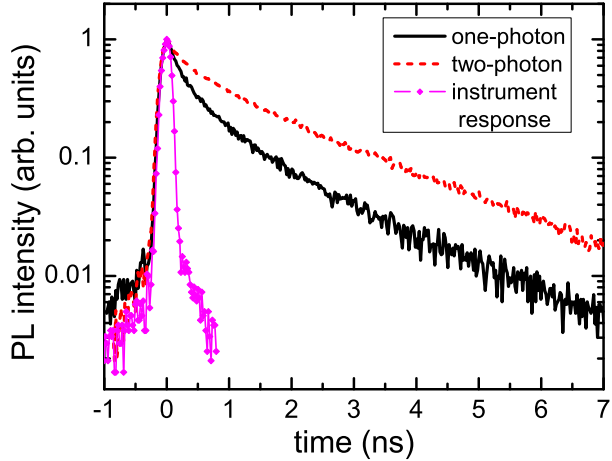}
\end{center}
\caption{\label{figTRPLannealed} (Color online) Time-resolved photoluminescence of annealed ZnO under one-photon and two-photon excitation.  The corresponding window of spectral integration was centered at the X-2LO transition ($3.177\eV\PM0.036\eV$).}
\end{figure}

The X-2LO exciton decays for annealed ZnO are shown in Fig.~\ref{figTRPLannealed} and quantified in Table~\ref{tabUVparams}.  The one-photon dynamics are nearly identical to the case of unannealed ZnO---a dominant fast decay component characterized by $\tau\sub{f}\SiM330\ps$, followed by a slow component with $\tau\sub{s}\SiM1.7\ns$.  On the sub-nanosecond time scale, nonradiative surface recombination still dominates because the annealing process has not significantly altered the density of surface traps.  In fact, the fast decay component is somewhat faster compared to that of unannealed ZnO because additional reabsorption is occurring near the surface [inset of Fig.~\ref{figSRTRPLtrt}(a)].

The annealed ZnO decay is noticeably different for the case of two-photon excitation, where the fast component is less dominant ($A\sub{f}\SiM0.5$) and the slow decay is longer ($\tau\sub{s}\GtR2\ns$).  These results are consistent with the conclusions drawn from the time-integrated PL and QE data [Fig.~\ref{figSRTRPLtrt}(a)]:  The annealing process has reduced the density of \textit{bulk} nonradiative traps; the slow decay time associated with the exciton's radiative lifetime is therefore longer---closer to the value observed in high-quality single-crystal ZnO ($\tau\sub{s}\spr{bulk}\SiM3\ns$).  The annealing process has also created a high density of green-emitting defects whose presence correlates with increased reabsorption of near-band-edge luminescence.  For volume excitation of the annealed ZnO, the smaller value of $A\sub{f}$ suggests a reduction in nonradiative surface recombination:  Of the excitons created in the bulk of the material which then propagate toward the surface, a large fraction of them is absorbed by green defects before reaching the surface, where nonradiative surface recombination would have occurred (and in which case the value of $A\sub{f}$ would have been larger).

The time-resolved near-band-edge PL spectra for annealed ZnO are shown in Fig.~\ref{figSRTRPLtrt}(b)--(c).  As was the case for unannealed ZnO, the differences in one-photon and two-photon PL spectra are maintained on every time scale.  The lack of time-evolving spectral distributions again suggests that the PL differences are due to reabsorption and distinct surface/bulk nonradiative trap distributions acting upon the initial spatial distributions of carriers, rather than exciton trapping or surface--bulk diffusion determining the dynamics.

\subsection{\label{secAnnealedGR} Defect-related green luminescence}
\begin{figure}[]
\begin{center}
\includegraphics[width=8cm]{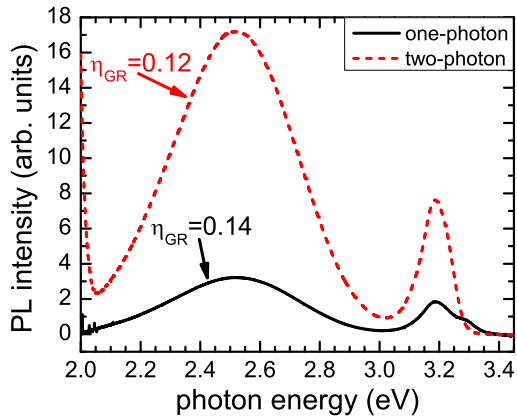}
\end{center}
\caption{\label{figGRPLtrt} (Color online) Time-integrated photoluminescence of annealed ZnO under one-photon and two-photon excitation.}
\end{figure}

Finally, we turn our attention to the broadband, defect-related green PL band shown in Fig.~\ref{figGRPLtrt}.  The spectrally integrated efficiency $\eta\sub{GR}$ is higher for surface excitation, which confirms our previous assertion that the majority of green-emitting defects are spatially concentrated in the near-surface region of the particle and are excited by one-photon excitation.  The green QE drops only slightly for volume excitation because the annealing process has reduced the density of bulk nonradiative recombination centers; the excitons are now being generated throughout the volume of the material, but most of them are able to propagate to the near-surface region and be absorbed by green traps without first undergoing nonradiative recombination in the bulk.  Thus, the small reduction in green QE for volume compared to surface excitation confirms that 1) the green-emitting traps are located primarily in the surface region of the particle and 2) the annealing process reduces the density of bulk nonradiative traps.  Because the green-emitting defects are concentrated near the sample surface, the main recombination channel competing with green PL is nonradiative surface recombination.

\subsection{\label{secModeling} Modeling}
A rigorous, quantitative model describing the results of Secs.~\ref{secUnannealed}--\ref{secAnnealedGR} is well beyond the scope of this experimental investigation.  Such a model must attempt to describe, among other factors, the spatial gradients of surface, bulk, and green-emitting defects.  None of these distributions can be inferred with high spatial resolution from the optical data presented here.  Even if these distributions were known, the convolution of these trapping effects with the geometry of the particles and the penetration depth of the exciting laser requires a complex model considering spatial, temporal, and spectral evolutions of the luminescence signals.  

However, for a qualitative understanding of how the various radiative and nonradiative recombination processes contribute to the observed quantum efficiencies, it is instructive to consider a simple, single-particle rate equation model for the exciton population.  It is assumed that excitons are created and thermalized on a time scale much shorter than the luminescence decays so that processes leading to the thermalized exciton population may be neglected.  This assumption is valid for the excitation fluence used to generate the data in Secs.~\ref{secUnannealed}--\ref{secAnnealedGR} because it corresponds to a low photogenerated carrier density:  A fluence of $3\uJcmsq$ creates an estimated carrier density of $\siM1\EE{18}\invvol$ if we take the laser penetration depth to be $\siM50\nm$.

In this section a series of rate equations will be considered in order of increasing complexity.  The complexity is related to the number of recombination channels competing with exciton luminescence, as inferred from the data in Secs.~\ref{secUnannealed}--\ref{secAnnealedGR} for different combinations of samples (unannealed versus annealed) and excitation (one-photon/surface versus two-photon/bulk).

\subsubsection{\label{subsecModelTPEannBulk} Two-photon excitation, annealed sample, bulk region}
First we consider an element of volume within the bulk region of the annealed ZnO sample.  The annealing process has drastically reduced the density of bulk nonradiative traps so that nonradiative recombination within this region will be ignored.  In the case of two-photon excitation, carriers are photogenerated uniformly throughout the volume element.  After initial generation of an exciton population from these photoexcited carriers, the only processes under consideration are depletion of excitons via radiative recombination and creation of excitons via reabsorption.  The exciton population $x$ is given by
\begin{equation*}\label{}
\dot{x}=+g-\frac{x}{\tau\sub{rad}}\,,\;x(t=0)=G\,.
\end{equation*}
The initial exciton population created by the laser is given by the initial condition $G$ (proportional to the laser fluence), and $\tau\sub{rad}$ is the intrinsic radiative exciton lifetime.  Exciton creation via reabsorption is described by the $g$ term.  We assume that the rate of exciton creation due to reabsorption is proportional to the existing exciton population according to an unknown factor $\tau\sub{reabs}^{-1}$:
\begin{equation*}\label{}
\dot{x}=+\left(\frac{1}{\tau\sub{reabs}}\right)x-\frac{x}{\tau\sub{rad}}=-\left(\frac{1}{\tau\sub{rad}}-\frac{1}{\tau\sub{reabs}}\right)x=-\frac{x}{\tau\sub{x}}\,,
\end{equation*}
where $\tau\sub{x}\EquiV(\tau\sub{rad}^{-1}-\tau\sub{reabs}^{-1})^{-1}$ is the effective exciton lifetime that takes into account reabsorption.  It is this effective lifetime that is measured by TRPL because the instantaneous UV PL intensity $I\sub{X}(t)$ is proportional to the exciton population, i.e., $I\sub{X}(t)\ProptO x(t)\ProptO \e{-t/\tau\sub{x}}$.~\cite{Bebb72}  We require $\tau\sub{reabs}^{-1}\LesS\tau\sub{rad}^{-1}$ so that the exciton population asymptotically decays with time rather than increases.  Therefore, reabsorption slows down the radiative decay process, but it does not change the single-exponential nature of the decay.  The effective lifetime $\tau\sub{x}$ can be approximated as the slow component of the UV TRPL data because this component corresponds to exciton luminescence originating in the bulk of the material.  For the case of two-photon excitation under consideration here, $\tau\sub{x}\ApproX2.05\ns$ (Table~\ref{tabUVparams}).

\subsubsection{\label{subsecModelTPEunannBulk} Two-photon excitation, unannealed sample, bulk region}
Now we consider the bulk region of the unannealed sample, where dynamics following two-photon excitation differ from the annealed sample through the presence of bulk nonradiative traps that introduce a parallel decay channel characterized by a trapping time $\tau\sub{nrb}$:
\begin{equation*}\label{}
\dot{x}=-\frac{x}{\tau\sub{x}}-\frac{x}{\tau\sub{nrb}}=-\left(\frac{1}{\tau\sub{x}}+\frac{1}{\tau\sub{nrb}}\right)x\,.
\end{equation*}
Again, the single-exponential nature of the decay is not altered; the decay is simply accelerated.  An estimate of $\tau\sub{nrb}$ may be obtained by considering the slow decay component in this sample (Table~\ref{tabUVparams}) to be the effective lifetime taking both decay channels into account:  $(1.50\ns)^{-1}\EQ\tau\sub{x}^{-1}\PluS\tau\sub{nrb}^{-1}$ implies that $\tau\sub{nrb}\ApproX5.59\ns$.  It is worth noting that this value is within an order of magnitude of an estimated nonradiative lifetime in ZnO from the literature.~\cite{Bylander78}

It is also interesting to consider the predicted UV quantum efficiency:  The internal quantum efficiency $\eta\sub{UV}\spr{int}$ is given by the competition between radiative and nonradiative decay rates,
\begin{equation*}\label{}
\eta\sub{UV}\spr{int}\equiv\frac{\tau\sub{x}^{-1}}{\tau\sub{x}^{-1}+\tau\sub{nrb}^{-1}}\approx 0.73\,.
\end{equation*}
The experimentally accessible \textit{external} quantum efficiency $\eta\sub{UV}\spr{ext}$ may be estimated by taking into account the probability $f\sub{esc}\spr{refl}$ of optical escape due to Fresnel effects [$f\sub{esc}\spr{refl}(\hw\sub{X-2LO}\EQ3.18\eV)$ in Fig.~\ref{figPLEopticalEscape}(b)]:
\begin{equation*}\label{eqEtaExt}
\eta\sub{UV}\spr{ext}=f\sub{esc}\spr{refl}\cdot\eta\sub{UV}\spr{int}\approx0.18\cdot\eta\sub{UV}\spr{int} \,\implies\, \eta\sub{UV}\spr{ext}\approx0.13\,.
\end{equation*}
This value is much larger than the UV quantum efficiencies reported in Secs.~\ref{secUnannealed}--\ref{secAnnealedUV} because nonradiative surface recombination has not yet been taken into account.  

\subsubsection{\label{subsecModelOPEunannSurface} One-photon excitation, unannealed sample, surface region}
Now we account for the effect of nonradiative \textit{surface} recombination by considering the fast decay component of the unannealed sample's UV PL following one-photon (near-surface) excitation.  Assuming the intrinsic exciton reabsorption and radiative decay processes are identical to those in the bulk region, there is now competition between the intrinsic exciton lifetime $\tau\sub{x}\EQ2.05\ns$ and a fast nonradiative surface recombination channel characterized by a trapping time $\tau\sub{nrs}$:
\begin{equation*}\label{}
\dot{x}=-\frac{x}{\tau\sub{x}}-\frac{x}{\tau\sub{nrs}}=-\left(\frac{1}{\tau\sub{x}}+\frac{1}{\tau\sub{nrs}}\right)x\,.
\end{equation*}
The effective lifetime near the surface for this sample and excitation condition is $0.40\ns$ (Table~\ref{tabUVparams}).  From this measured lifetime, the nonradiative surface trapping time may be estimated:  $(0.40\ns)^{-1}\EQ\tau\sub{x}^{-1}\PluS\tau\sub{nrs}^{-1}$ implies that $\tau\sub{nrs}\ApproX0.50\ns$.  Again, this value is reasonable based on the approximate quantum efficiency it predicts,
\begin{equation}\label{eqEtaInt}
\eta\sub{UV}\spr{int}=\frac{\tau\sub{x}^{-1}}{\tau\sub{x}^{-1}+\tau\sub{nrs}^{-1}}\approx 0.20\,\implies\,\eta\sub{UV}\spr{ext}\approx0.036\,.
\end{equation}
Now that nonradiative surface recombination has been considered, the estimated UV external quantum efficiency is within a factor of 3 of the experimentally measured $\eta\sub{UV}\EQ0.014$.  An even more accurate estimate would result if bulk nonradiative recombination were simultaneously considered.

\subsubsection{\label{subsecModelOPEannSurface} One-photon excitation, annealed sample, surface region}
Finally, we consider the near-surface region of the annealed sample.  The fast decay component under one-photon excitation is accelerated compared to the unannealed sample because excitons are additionally able to decay through green-emitting defects in the annealed sample.  The trapping of excitons at green-emitting defects, as characterized by a trapping time $\tau\sub{gr}$, competes with both exciton trapping at nonradiative surface traps and intrinsic radiative recombination:
\begin{equation*}\label{eqModelAllChannels}
\dot{x}=-\frac{x}{\tau\sub{x}}-\frac{x}{\tau\sub{nrs}}-\frac{x}{\tau\sub{gr}}=-\left(\frac{1}{\tau\sub{x}}+\frac{1}{\tau\sub{nrs}}+\frac{1}{\tau\sub{gr}}\right)x\,.
\end{equation*}
The values $\tau\sub{x}\EQ2.05\ns$ and $\tau\sub{nrs}\EQ0.50\ns$ may be used to estimate $\tau\sub{gr}$, again using the appropriate overall lifetime from Table~\ref{tabUVparams}:  $(0.33\ns)^{-1}\EQ\tau\sub{x}^{-1}\PluS\tau\sub{nrs}^{-1}\PluS\tau\sub{gr}^{-1}$ implies that $\tau\sub{gr}\ApproX2.0\ns$.  The green defect trapping time is almost identical to the intrinsic exciton lifetime, which suggests that excitation of green-emitting defects occurs only during the lifetime of the excitons.  This result supports the claim that the green defect excitation mechanism is strongly correlated with the presence of excitons.

Using an equation analogous to Eq.~\eqref{eqEtaInt}, it is readily apparent that the predicted UV quantum efficiency decreases with the addition of green-emitting defects.  A decrease in the one-photon QE is indeed observed experimentally when comparing the annealed sample to the unannealed sample.  Moreover, this simple rate equation model predicts the \textit{green} quantum efficiency with surprising accuracy:
\begin{equation*}\label{eqModelGRQE}
\eta\sub{GR}\spr{int}=\eta\sub{GR}\spr{ext}=\frac{\tau\sub{gr}^{-1}}{\tau\sub{x}^{-1}+\tau\sub{nrs}^{-1}+\tau\sub{gr}^{-1}}\approx 0.17\,,
\end{equation*}
which compares with the experimental value $\eta\sub{GR}\EQ0.14$ [Fig.~\ref{figSRTRPLtrt}(a)].  No reduction in green QE due to Fresnel reflections and reabsorption is required because there are no nonradiative traps that can absorb green emission; consequently, all of the emitted green photons eventually escape the sample.

\section{\label{secSummary} Summary}
The influence and spatial distributions of radiative and nonradiative traps in unannealed, nanometer-scale commercial ZnO particles and their annealed, micrometer-scale modified forms have been characterized optically using femtosecond one-photon and two-photon excitation.  The unannealed particles exhibit primarily near-band-edge (ultraviolet) emission with a quantum efficiency of $\eta\sub{UV}\LesssiM1\%$.  A comparison of one-photon and two-photon QEs and (TR)PL data suggests that high densities of both surface and bulk nonradiative traps reduce the emission efficiency from its theoretical upper limit of $\gtrsiM 10\%$.  The annealed particles additionally exhibit a broad, visible-wavelength band that is characteristic of ZnO:Zn phosphors.  This emission band, which exhibits a relatively large quantum efficiency ($\eta\sub{GR}\GtR10\%$) and decays on a microsecond time scale, originates from defects located primarily near the surface of the particles.  The annealing process also appears to reduce the density of bulk nonradiative traps.  A simple rate equation model is developed to ascertain qualitatively the effects of the various radiative and nonradiative recombination processes on the measured quantum efficiencies.

\begin{acknowledgments}
This work was supported by RDECOM/AMRDEC Competitive In-House Laboratory Independent Research (ILIR) funding.  JVF and HOE thank \"{U}. \"{O}zg\"{u}r for a critical reading of an early version of this manuscript.
\end{acknowledgments}
%


\begin{thebibliography}{36}
\bibitem[{FNs()}]{FNsolvent}
\bibinfo{note}{A portion of the powder was alternatively pressed between two
  fused silica plates to confirm that the solvent had no effect on the optical
  properties of the ZnO.}
\bibitem[{FNa()}]{FNabsorb}
\bibinfo{note}{The number of photons absorbed by the sample is measured as a
  part of the quantum efficiency measurement procedure.}
\bibitem[{FNl()}]{FNltple}
\bibinfo{note}{The details of the PLE traces (fine structure, peak locations,
  etc.) are beyond the scope of the current discussion and are the subject of a
  more detailed future investigation.}
\bibitem[{FNr()}]{FNrtple}
\bibinfo{note}{Room-temperature PLE was also performed on the X-2LO emission
  peak near the band edge, and the excitation profile was virtually identical
  to that shown in Fig.~\ref{figPLEopticalEscape}(a), except the
  ``enhancement'' at $3.26\eV$ was not as pronounced.}
\bibitem[{FNn()}]{FNnormalization}
\bibinfo{note}{The normalization factor of $2\pi$ is obtained by performing the
  integrals for the case $n\EQ1$ ($\theta\sub{c}\EQ\pi/2$).}
\end{thebibliography}

\end{document}